# Accurate GW frontier orbital energies of 134 kilo molecules

Artem Fediai[1,3,*], Patrick Reiser[1], Jorge Enrique Olivares Peña[1], Pascal Friederich[1,2], Wolfgang Wenzel[1]

[1] *Institute of Nanotechnology, Karlsruhe Institute of Technology, Hermann-von-Helmholtz-Platz 1, 76344 Eggenstein-Leopoldshafen, Germany*

[2] *Institute of Theoretical Informatics, Karlsruhe Institute of Technology, Am Fasanengarten 5, 76131 Karlsruhe, Germany*

[3] *Current address: Nanomatch GmbH, Griesbachstraße 5, 76185 Karlsruhe, Germany*

\* Corresponding author: artem.fediai@nanomatch.com

## Abstract

The QM9 dataset [Scientific Data, Vol. 1, 140022 (2014)] became a standard dataset to benchmark machine learning methods, especially on molecular graphs. It contains geometries as well as multiple computed molecular properties of 133,885 compounds at B3LYP/6-31G(2df,p) level of theory, including frontier orbitals (HOMO and LUMO) energies. However, the accuracy of HOMO/LUMO predictions from density functional theory, including hybrid methods such as B3LYP, is limited for many applications. In contrast, the GW method significantly improves HOMO/LUMO prediction accuracy, with mean unsigned errors in the GW100 benchmark dataset of 100 meV. In this work, we present a new dataset of HOMO/LUMO energies for the QM9 compounds, computed using the GW method. This database may serve as a benchmark of HOMO/LUMO prediction, delta-learning, and transfer learning, particularly for larger molecules where GW is the most accurate but still numerically feasible method. We expect this dataset to enable the development of more accurate machine learning models for predicting molecular properties

## Background and summary

The availability of a large datasets of sufficiently *accurate* values of frontier orbital energies (i.e., highest occupied and lowest unoccupied orbitals, HOMO and LUMO, respectively) or rather ionization energies (ionization potential and electron affinity, IE and EA, respectively) is a prerequisite for the virtual design of molecules using data-driven, in particular machine learning based, approaches. Virtual materials design is relevant for many applications, ranging from organic electronics[1,2], functional materials[3] and thermo-electrics[4] to homogeneous catalysis[5].
An ubiquitous method suitable to compute IP and EA in the course of high-throughput screening is density functional theory (DFT)[6]. In DFT, the many-body system of interacting electrons is replaced with a system of non-interacting quasi-particles in the field of the exchange-correlation potential ($V_{xc}[n]$), which is a unique functional of the electron density $n$. Although exact in theory, practical DFT requires severe approximations of $V_{xc}[n]$, which can be represented as a chain of progressively more accurate (and more expensive) approximations called Jacob's ladder[7]. Its first rungs, local density approximation (LDA) and generalized gradient approximation (GGA) are the most widely used approximations. It is well known, however, that these approximations systematically underestimate fundamental HOMO-LUMO gaps by up to 5 eV[8]. Unfortunately, neither the highest implemented rungs of Jacob's ladder[9], nor empirical functionals, nor hybrid functionals can closely approach chemical accuracy (1 kcal / mole = 0.0434 eV)[10].

In contrast to DFT, the GW method allows to systematically increase the accuracy of computing single-particle excitation spectra (including EA and IP) by eliminating some critical problems of DFT, e.g. the interpretation of HOMO and LUMO quasi-particle energies as -IP and -EA, which is an assumption that does not hold in all general cases[11,12]. According to recent reports[13–15], GW accuracy on various test sets reaches 0.1(0.2) eV, a factor of 2(4) larger than the chemical accuracy.

Here we use the non-self-consistent GW ($G_0W_0$) and eigen-value-self-consistent GW (denoted as GW) based on GGA DFT (namely the PBE exchange-correlation functional[16]) as an initial guess for GW. These two methods are later denoted as $G_0W_0$@PBE and GW@PBE, respectively. A discussion on theoretical details of the GW method can be found in the Supporting Information. Our data includes HOMO/LUMO and IP/EA energies computed at various levels of theory, ranging from GGA DFT with aug-cc-DZVP basis set to self-consistent GW@PBE extrapolated to the basis set limit. We explain the structure of the dataset, and analyze as well as compare the distribution of energy levels across various levels of theory. Finally, the quality of the basis set limit scheme is analyzed, and results obtained from the quantum chemistry package CP2K[17] are compared to Gaussian 09 calculations[18]. Notably, this dataset represents the largest collection of GW simulations reported in literature to date. While the accuracy of the method used to compute HOMO/LUMO in original QM9 dataset[19] is low when compared to experimental results, our reported GW IP/EA energies can be used for machine learning methods that are aimed at accurately predicting ionization energies of small molecules.

## Methods

We employ CP2K Gaussian Augmented Plane Wave (GAPW) method for both DFT and GW simulations. DFT total energies convergence criterion is $10^{-6}$ Hartree. Realspace grids settings: The cutoff of the finest grid level (CUTOFF) is 500 Ry, the number of multigrids (NGRIDS) is 5; the relative cutoff (REL_CUTOFF) is set to 50 Ry. The simulation cell size (ABC) is set to be 10 Angstroms larger than the linear size of the molecule.

GW simulations were performed using 50 quadrature points (QUADRATURE_POINTS) in resolution-of-identity Random Phase Approximation (RI-RPA) as a default value, crossing search (CROSSING_SEARCH) is set to NEWTON. These simulations converged for about 99% of all molecules (132,151 molecules of 133,885). If the self-consistent quasiparticle solutions were not found within the iteration limit of 20 or the GW algorithm returned NaN values (manifestation of the instability issues) settings were changed: (1) more quadrature points were set: 100, 200, or 500; (2) CROSSING_SEARCH is set to BISECTION instead of NEWTON; (3) if this did not lead to convergence, CUTOFF/REL_CUTOFF was increased to

1000/50, respectively; (4) at last, the Fermi level offset (FERMI_LEVEL_OFFSET) with a default value of 0.02 Hartree set to 0.04 Hartree. As a result, 1351/150/233 molecules converged with 100/200/500 QUADRATURE_POINTS. An example of the default input file for molecule 123456 of the dataset is provided in Supporting Information.

## Data records

HOMO and LUMO levels of the whole QM9 dataset molecules were computed in this work using the correlation-consistent basis set aug-cc-DZVP[20] and the PBE functional[16] followed by eigenvalue self-consistent GW calculations as implemented in CP2K[21], which takes the PBE solution as an initial guess (GW@PBE). The same procedure has been repeated for the aug-cc-TZVP basis set. With the GW results from two basis sets we extrapolate the energy to the infinite basis set limit, assuming that the energy is proportional to 1/*N* with *N* being the number of the basis functions[21]. We report HOMO/LUMO energies computed at the level of PBE, $G_0W_0$, GW, each with the two mentioned basis sets together with the corresponding extrapolated values. The notation and dataset labels for HOMO and LUMO orbital energies as computed with DFT as well as GW are summarized in **Table 1**.

Although the extrapolation to the basis set limit at the PBE level was performed, it was not actually necessary as the convergence was essentially reached at the level of the aug-cc-DZVP basis set. However, it should be noted that GW HOMO/LUMO energies exhibit slower basis-set convergence[21], and the extrapolation is essential to attain the nominal GW accuracy.

*Table 1. Notations used for orbital / quasiparticle energies.*

| Notation in the manuscript | Notation in database files | Meaning | Level of theory |
|---|---|---|---|
| $\epsilon_{\text{HOMO}}^{\text{DFT}}$ | homo* | HOMO energy computed using PBE functional | Basis sets: aug-cc-DZVP and aug-cc-TZVP *extrapolated* to the basis set limit. |
| $\epsilon_{\text{LUMO}}^{\text{DFT}}$ | lumo | LUMO energy computed using PBE functional | Basis sets: the same as above |
| $\epsilon_{\text{HOMO}}^{\text{GW}}$ | occ_scf | GW quasiparticle energy of the HOMO computed self-consistently with the PBE starting guess | Basis sets: the same as above GW: quasiparticle eigenvalue-only self-consistent with PBE as an initial guess |
| $\epsilon_{\text{LUMO}}^{\text{GW}}$ | vir_scf | GW quasiparticle energy of the LUMO computed self-consistently with the PBE starting guess | Basis sets and GW: as above |
| $\epsilon_{\text{HOMO}}^{G_0W_0}$ | occ_0 | $G_0W_0$ quasiparticle energy of the HOMO with the PBE starting guess | Basis sets: the same as above GW: "one-shot" GW with PBE initial guess (not self-consistent). |
| $\epsilon_{\text{LUMO}}^{G_0W_0}$ | vir_0 | $G_0W_0$ quasiparticle energy of the LUMO with the PBE starting guess | Basis sets and GW: the same as above |
| $\tilde{\epsilon}_{\text{HOMO}}^{G_0W_0}$ | occ | $G_0W_0$ quasiparticle energy of the HOMO with the PBE starting guess, assuming the HOMO at PBE remains HOMO at $G_0W_0$ level (not, for instance, HOMO-1) | Basis sets: the same as above GW: "one-shot" GW with PBE initial guess (not self-consistent). |
| $\tilde{\epsilon}_{\text{LUMO}}^{G_0W_0}$ | vir | $G_0W_0$ quasiparticle energy of the LUMO with the PBE starting guess, assuming the LUMO at PBE remains LUMO at $G_0W_0$ level (not, for instance, HOMO+1) | Basis sets and GW: as above |
| $\epsilon_{\text{<orbital>}}^{\text{<method>}}(2)$, $\epsilon_{\text{<orbital>}}^{\text{<method>}}(3)$, $\epsilon_{\text{<orbital>}}^{\text{<method>}}(4)$ | <name>(2), <name>(3), <name>(4) where <name> is one of the notations from above, e.g.: homos(2) is $\epsilon_{\text{HOMO}}^{\text{DFT}}(2)$ | Energies, computed for a specific basis set. Method depends on <orbital> and <method>. Possible values: <orbital>: HOMO or LUMO <method>: DFT or GW | Basis set: (2): aug-cc-DZVP (3): aug-cc-TZVP (4): aug-cc-QZVP |

* two extrapolation methods are used to obtain energy levels in the infinite basis set limit. Method 1: $\sim 1/n$, $n$ being the number of basis functions. Method 2: $\sim 1/N^3$ with $N$ being the basic set size (*i.e.*, DZ: 2, TZ: 3, QZ: 4). They are saved as a list, [<method 1>, <method 2>]. Assumptions of method 1 are found to be empirically better, thus it is used throughout the paper.

### *Orbital and quasiparticles energies in the basis set limit*

**Figure 1** shows the distribution of the PBE and GW HOMO/LUMO energies in the infinite basis set limit. The obtained HOMO position depends on the level of the theory. The systematic difference between PBE and GW level of theory is considerable: DFT with the PBE functional yields a mean HOMO energy of -5.79 eV, while $G_0W_0$@PBE yields a mean HOMO energy of -9.02 eV, which is approximately 3.2 eV lower. GW@PBE is on average approximately 0.9 eV lower than $G_0W_0$@PBE and yields a mean HOMO energy of -9.91 eV.

Noticeable is the difference between the distribution of $\epsilon_{\text{LUMO}}^{G_0W_0}$ and $\tilde{\epsilon}_{\text{LUMO}}^{G_0W_0}$ in the energy range between 1 eV and 1.5 eV. This means that many molecules with positive LUMO energy change the order of orbitals. Almost no such effect can be observed for the HOMO energy distributions.

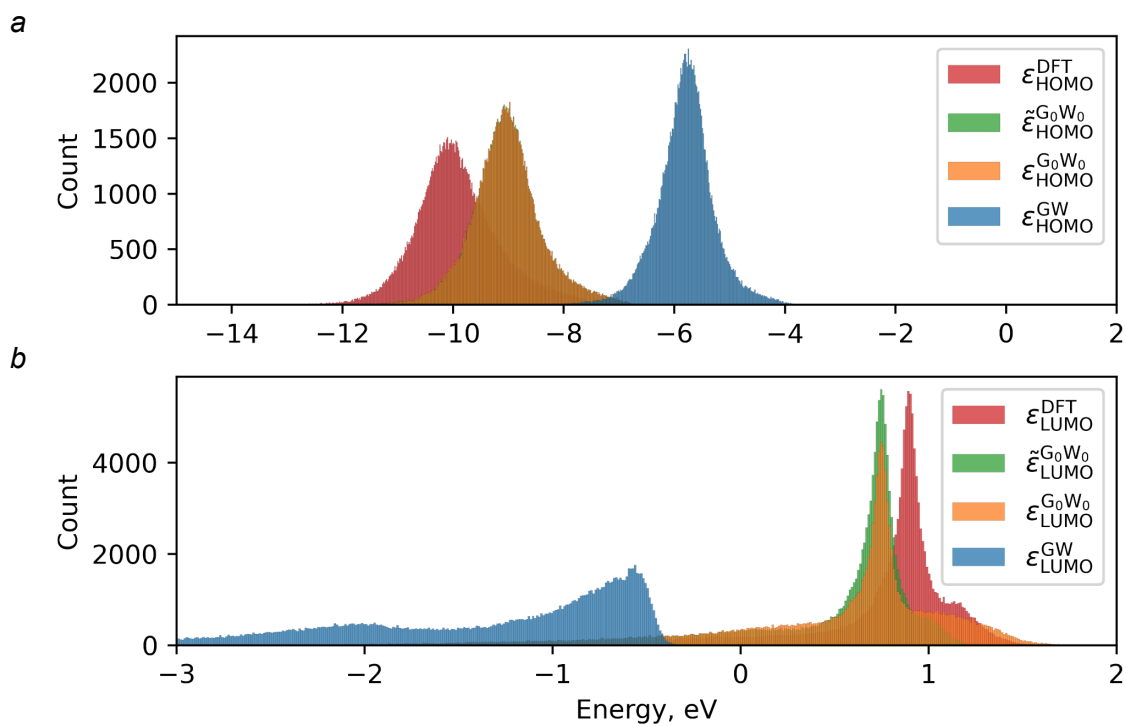

*Figure 1*. Distribution of HOMO (a) and LUMO (b) energies computed at various levels of theory from DFT to self-consistent GW. For notations see Table 1.

**Figure 2** shows the correlation of GW quasiparticle energies to corresponding DFT orbitals energies. While a few electron-volts difference between DFT and GW methods was obvious from **Figure 1**, linear regression fits in **Figure 2** show that the difference between GW and DFT contains large molecule-specific components. For instance, the average difference between $\epsilon_{\text{LUMO}}^{\text{GW}}$ and $\epsilon_{\text{LUMO}}^{\text{DFT}}$ depends on the orbital energy: it increases as $\epsilon_{\text{LUMO}}^{\text{DFT}}$ decreases (the slope of the dotted regression line in **Figure 2** is 0.48). Additionally, there is a large spread of the data (the mean absolute deviation of $\epsilon_{\text{LUMO}}^{\text{GW}}$ distribution is 0.34 eV). DFT HOMO energies correlate better to GW HOMOs than LUMO levels, e.g. for HOMOs, the coefficients of determination $R^2$ are 0.79 and 0.90 for GW and $G_0W_0$, whereas for LUMOs $R^2$ are 0.61 and 0.77 for GW and $G_0W_0$, respectively.

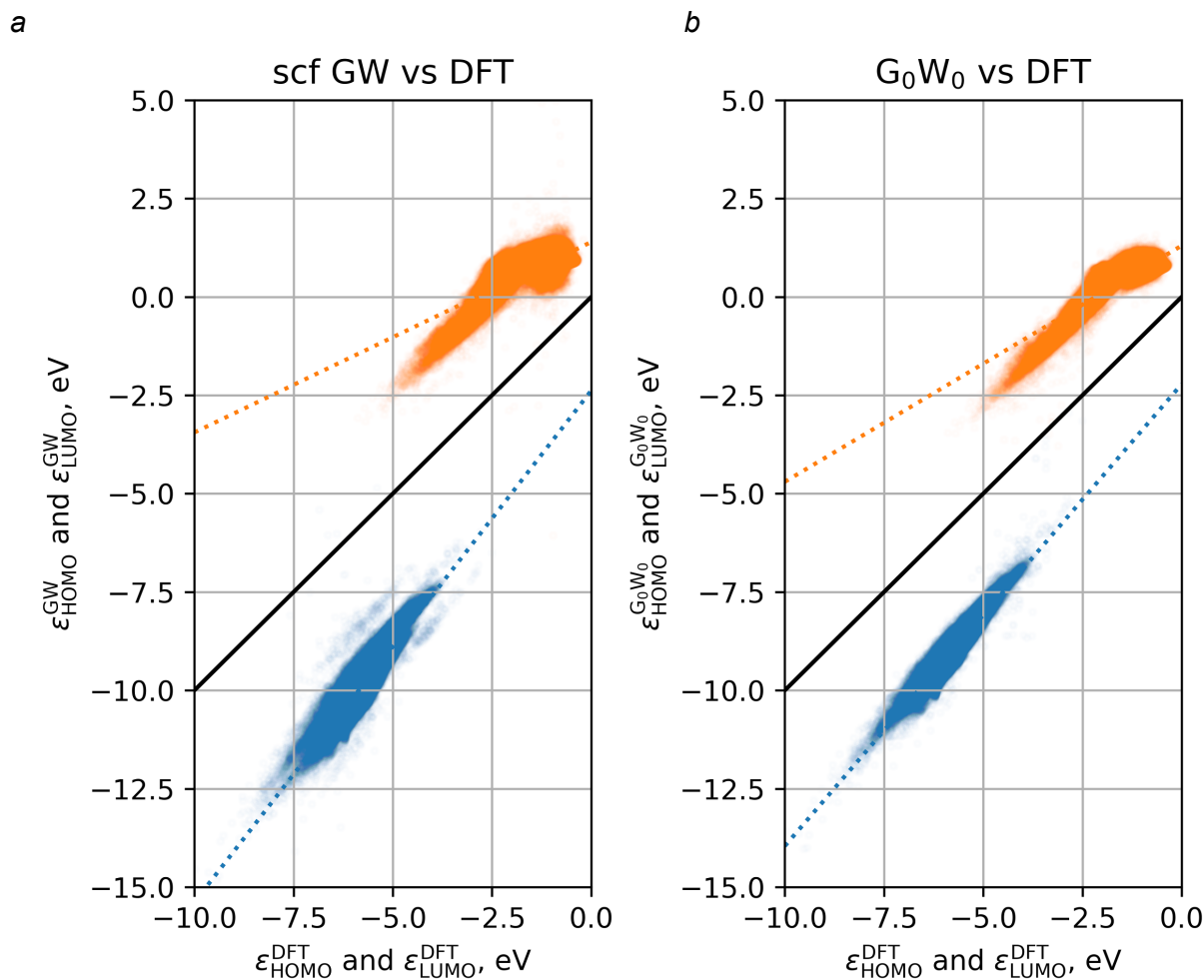

*Figure 2. Pair correlation plots of frontier orbitals as computed with GW and DFT methods: a) eigenvalue self-consistent GW vs. DFT, b) "one-shot" GW ($G_0W_0$) vs. DFT.*

Technical validation

***Benchmarking and choosing basis set limit extrapolation schemes***

Due to the slow basis set convergence of quasiparticle HOMO and LUMO energies in GW calculations, extrapolation to the complete basis set limits was carried out. GW energies of all QM9 molecules were computed using two all-electron basis sets of a different size: aug-cc-DZVP and aug-cc-TZVP, and then extrapolated using two basis set extrapolation schemes [13]. *Scheme 1* employs a linear fit on the HOMO or LUMO values versus the inverse cardinal number of the basis set $N_{basis}$ (GW HOMO/LUMO energy is assumed to be proportional to $1/N_{basis}$). *Scheme 2* extrapolates HOMO/LUMO energies against $1/N_{card}^3$ where $N_{card}$ is the cardinal number of the basis set (for example 2 for aud-cc-DZVP, 3 for aud-cc-QZVP, etc.).

To test the quality of the extrapolation from these two relatively smaller aug-cc basis sets, one hundred pseudo-random molecules from the QM9 dataset were simulated with the larger aug-cc-QZVP basis set.

The extrapolated GW HOMO and LUMO energies analyzed in this paper is based on *Scheme* 1, although the data set contains extrapolated values for both *Scheme* 1 and *Scheme* 2. For Scheme 1, the smallest mean absolute error (mae) is reached for $\epsilon_{\text{HOMO}}^{G_0W_0}$ of 6.0 meV, more than an order of magnitude more than the GW method accuracy. The worst extrapolation quality is observed for $\epsilon_{\text{LUMO}}^{\text{GW}}$ with a mae of 37.0 meV. However, this is still acceptable, as it is a few times smaller than the GW mean error (around 100…200 meV [13]). The extrapolation errors are defined as the normalized sum of the absolute differences of the extrapolated values computed with the use of two (aug-cc-DZVP, aug-cc-TZVP) and three (aug-cc-DZVP, aug-cc-TZVP, and aug-cc-QZVP) basis sets:

$$\text{mae}_{<\text{orbital}>}^{<\text{method}>} = \frac{1}{N_{\text{mol}}} \sum_i |\epsilon_{<\text{orbital}>,i}^{<\text{method}>}(2,3,4) - \epsilon_{<\text{orbital}>,i}^{<\text{method}>}(2,3)|$$

where $<\text{method}>$ is either GW or $G_0W_0$, $<\text{orbital}>$ is either HOMO or LUMO, $i$ is the molecular index, $N_{\text{mol}}$ is the number of molecules, which is 100. $\epsilon_{<\text{orbital}>,i}^{<\text{method}>}(2,3,4)$ and $\epsilon_{<\text{orbital}>,i}^{<\text{method}>}(2,3)$ denote extrapolated energies computed using three and two basis sets, respectively. $\epsilon_{<\text{orbital}>,i}^{<\text{method}>}(2,3)$ is identical to $\epsilon_{<\text{orbital}>,i}^{<\text{method}>}$, and is added here for clarity.

Unfortunately, the overall acceptable mean absolute error magnitude is accompanied with a few outliers (see Figure 3), which are much more pronounced for LUMO than HOMO extrapolation errors. The spread of LUMO in **Figure 2**, therefore, may potentially be larger than that of HOMOs due to extrapolation errors, not physical reasons.

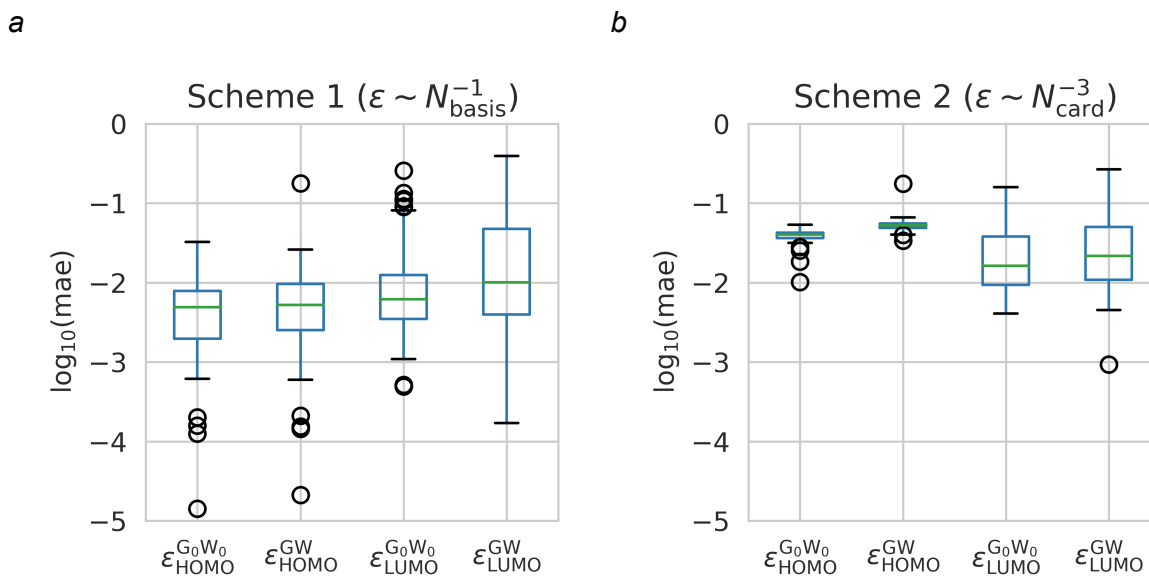

*Figure 3*. Visualization of the extrapolation errors of HOMO/LUMO computed at GW@PBE and G0W0@PBE levels: a) Scheme 1, b) Scheme 2. The extrapolation errors are computed for 100 random molecules from the QM9 dataset. They are defined as the normalized sum of the absolute differences of the extrapolated values computed with the use of two (aug-cc-DZVP, aug-cc-TZVP) and three (aug-cc-DZVP, aug-cc-TZVP, and aug-cc-QZVP) basis sets. Scheme 1 is up to one order of magnitude more accurate than Scheme 2.

### *Benchmark calculations using B3LYP.*

Original simulations of HOMO and LUMO energies in the QM9 data set were performed using the B3LYP functional and a 6-31G(2df,p) basis set using the Gaussian 09 software [Frisch, M. J. et al. *Gaussian 09, Revision d.01* (Gaussian, Inc., 2009).] [18]. In addition to the aforementioned computational protocol for DFT/GW simulations, we also performed B3LYP/6-31G(2df,p) calculations to estimate differences between CP2K[21] used here and the original work (Gaussian 09). Results are shown in **Figure 4a** for 100 randomly selected molecules from the QM9 dataset. While perfect correlation is observed for HOMOs (standard deviation of the HOMO differences is 11 meV), LUMO values demonstrate worse correlation (standard deviation of the LUMO differences is 73 eV). For LUMOs which have energies exceeding 1 eV, the orbital energies computed in this work are systematically lower than the original QM9 energy, which could be due to the fact that CP2K uses mixed localized/plane-wave basis sets to represent electron density, which is different in Gaussian.

### *Benchmark calculations for GW100 dataset*

The *GW*100 [13] dataset is a dataset of small molecules used to benchmark GW implementation in various quantum chemistry codes. The GitHub repository [22] contains, among others, HOMO quasiparticles energies computed using CP2K self-consistently at GW@PBE level using def2-QZVP basis set [23]. **Figure 4b** compares the organic molecules within GW100 with CP2K simulations at the same theory level. However, the exact equivalence of all computational settings cannot be assured as the full CP2K input files are not available. Apart from the outlier molecule Carbon tetrafluoride, named 75-73-0 in GW100 data repository (for which the error is 71 meV), the observed differences are small, with a mean unsigned error of 28 meV (including the outlier), which is substantially smaller than the accuracy of the GW method itself.

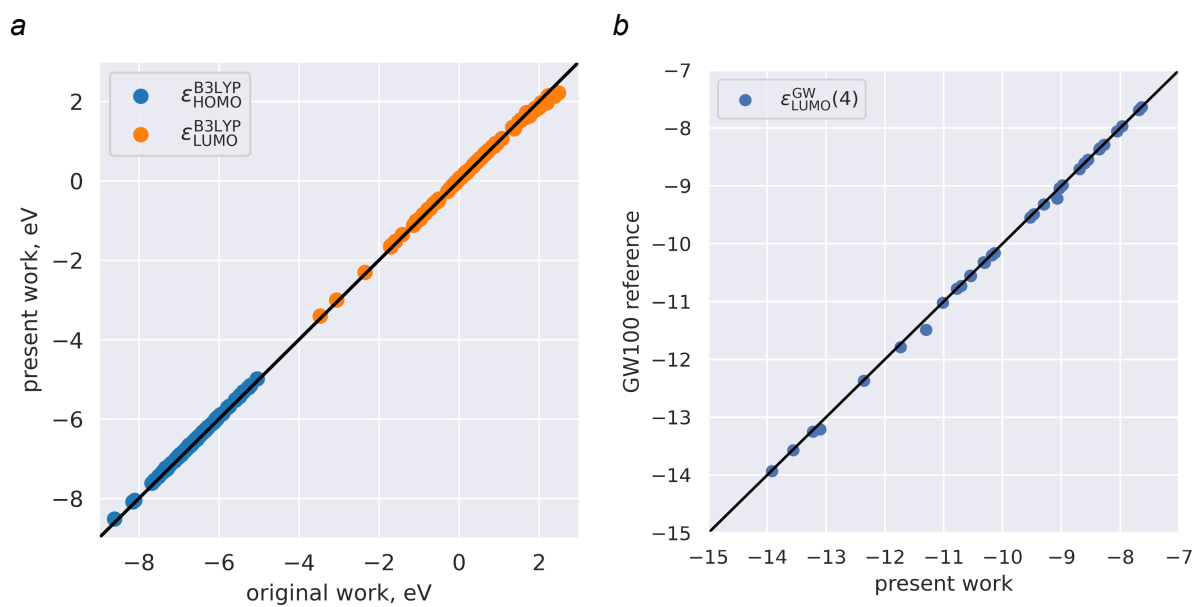

**Figure 4.** Benchmarking calculations: *a*) Correlation plot of HOMO and LUMO, contained in the original dataset (Gaussian 09) and here (CP2K). Theory level: B3LYP/6-31G(2df,p). *b*) Correlation plot of GW@PBE HOMOs, as deposited in the GW100 data set [22] in comparison with the present work (CP2K). Theory level: Self-consistent GW@PBE with a def2-QZVP basis set.

## *Computational resources and scaling*

Overall, it took 7,439,925 cpu hours to perform DFT and GW simulations in order to generate the scientific data reported. The total cpu time to make DFT and GW simulations for one molecule scales as $n_e^3$ with $n_e$ being the number of electrons of the molecule (see Figure 5). More details are visualized in **Supplementary Figure S1**, including distribution of computational time splitted by the different cpu model specifications. Hardware specifications used in this work are listed in **Supplementary Table 1**.

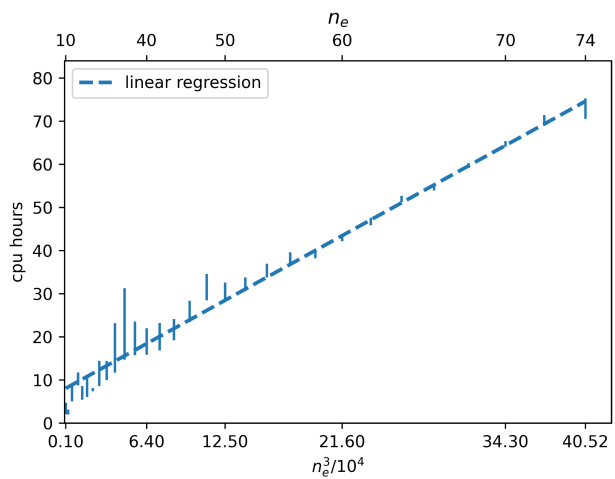

**Figure 5**. Scaling of the computation time (*cpu hours*) depending on the cubic number of electrons in a molecule, $n_e^3/10^4$. The upper horizontal axis is nonlinear, and represents the number of electrons $n_e$. Cubic cpu time scaling ($O(N^3)$) is observed for GW implementations.

## Usage notes

We presented accurate values of HOMO and LUMO of 134 kilo molecules, computed with an eigenvalue self-consistent GW method in a basis set limit, along with auxiliary data: $G_0W_0$, and DFT values of HOMO and LUMO orbitals. This data can be used to benchmark machine-learning methods, which aim at the accurate prediction of single-particle excitation energies. It contains many more molecules than the standard GW100 data set, and thus can also be used to benchmark new and existing GW codes.

## Code availability

An input file for the CP2K calculations can be found in the Supplementary Information. Further code is not required to reproduce the data presented in this article.

## Acknowledgement

The authors acknowledge support by the state of Baden-Württemberg through bwHPC and the German Research Foundation (DFG) through grant no INST 40/575-1 FUGG (JUSTUS 2 cluster); by the state of Baden-Württemberg through bwHPC and DFG through grant INST 35/1134-1 FUGG (MLS-WISO cluster). The authors acknowledge support by the state of Baden-Württemberg through bwHPC. P.F. acknowledges funding by ZIM project KK5139001APO.

# Supporting Information

# GW frontier orbital energies of 134 kilo molecules

Artem Fediai, Patrick Reiser, Jorge Enrique Olivares Peña, Pascal Friederich, Wolfgang Wenzel

# Short theory review of GW

The GW method[1–3] provides an approximation to the real many-body excitation spectra using single quasiparticle Green's functions $G$[4]. The effects of exchange (Pauli repulsion) and correlation (screening) are taken into account through an energy dependent self-energy $\Sigma_{GW}(E) = \Sigma_x + \Sigma_c(E)$ that depends on the Green's function. The Green's function enters a Dyson type of equation connecting the non-interacting (Hartree type) Green's function $G_H(z)$ with the full interacting Green's function: $G(z) = G_H(z) + G_H(z)\Sigma_{GW}(z)G(z)$, where z is a complex number. In the GW approximation, the vertex corrections as defined by Hedin[1] are ignored and therefore one can calculate the self-energy using the second Hedin's equation with a dynamically screened Coulomb potential $W$ which we write symbolically as : $\Sigma = iGW$. The GW approximation can be viewed as the generalization of the Hartree-Fock method where exchange energy is represented as $V_x = iGV$ with $V_x$ being a bare interaction potential. Bare and screened potentials differ by dielectric function $\epsilon$: $W = \epsilon^{-1}V$. The self-energy includes the effects of the static and dynamic screening (correlations) missing in the Hartree-Fock method and "underrepresented" in the practical DFT implementations.

Practical solution for the quasiparticles energy levels in GW is a self-consistent problem (thus another name for this GW implementation is scGW), just like the solution of the Kohn-Sham equations. This self-consistent solution ( solving Hedin's equations with an initial guess on the Green's function) does not depend on the initial choice of the reference system which can be taken from DFT or Hartree-Fock methods[5]. The problem is that the "full" solution is computationally demanding and the computation cost of a canonical GW growth as $O(N^6)$ with the number of particles. In practice, the so-called "full" self-consistent GW is rarely used. A "one-shot" GW, called $G_0W_0$ , is the least expensive scheme, where the self-energy $\Sigma$ is calculated once using the Green's function obtained with the initial guess (DFT in many cases) $G_{DFT}$ , in other words, just one iteration of the scGW is performed. In contrast to the full $GW$ method, $G_0W_0$ does depend on the initial guess of the reference system, thus, excellent DFT convergence is a must to obtain reliable results. There exist other flavors of GW which are sometimes called "partial self-consistent"[5] approaches like the so called quasiparticle-self-consistent GW (qsGW)  and the  quasiparticle eigenvalue-only self consistent GW (evGW). Although one expects the accuracy of  the self-consistent methods to be be higher than that of $G_0W_0$, $G_0W_0$ may be fairly accurate if using a reasonable initial guess, and outperforms scGW (in terms of memory and computational time resources) when screening a large amount of molecules. qs-GW and ev-GW lead to very similar results.

# Example of cp2k input file

```
&GLOBAL
  EXTENDED_FFT_LENGTHS TRUE
  PRINT_LEVEL LOW
  PROJECT_NAME 2
  RUN_TYPE ENERGY
&END GLOBAL
&FORCE_EVAL
  METHOD QUICKSTEP
  &DFT
      POTENTIAL_FILE_NAME POTENTIAL
      UKS FALSE
      MULTIPLICITY 1
      CHARGE 0
      BASIS_SET_FILE_NAME BASIS_CC_AUG_RI_NEW
      &SCF
      MAX_SCF 100
      EPS_SCF 1e-06
      SCF_GUESS RESTART
      ADDED_MOS 1000
      &DIAGONALIZATION T
      &END DIAGONALIZATION
      &MIXING
      METHOD BROYDEN_MIXING
      ALPHA 0.2
      BETA 1.5
      NBUFFER 8
      &END MIXING
      &END SCF
      &QS
      EPS_DEFAULT 1e-10
      EPS_PGF_ORB 1e-05
      METHOD GAPW
      &END QS
      &MGRID
      NGRIDS 5
      CUTOFF 500
      REL_CUTOFF 50
      &END MGRID
      &XC
      &XC_FUNCTIONAL PBE
      &END XC_FUNCTIONAL
      &WF_CORRELATION
      METHOD RI_RPA_GPW
      MEMORY 4000
      GROUP_SIZE 1
      ERI_METHOD OS
      &RI_RPA
      QUADRATURE_POINTS 50
      SIZE_FREQ_INTEG_GROUP -1
      RI_G0W0 TRUE
      &RI_G0W0
          CORR_MOS_OCC 20
          CORR_MOS_VIRT 20
          CROSSING_SEARCH NEWTON
          EV_SC_ITER 20
          RI_SIGMA_X .TRUE.
```

```
                ANALYTIC_CONTINUATION PADE
            &END RI_G0W0
            &HF
                FRACTION 1.0
                &SCREENING
                    EPS_SCHWARZ 1e-11
                    SCREEN_ON_INITIAL_P FALSE
                &END SCREENING
                &MEMORY
                    MAX_MEMORY 500
                &END MEMORY
            &END HF
          &END RI_RPA
        &END WF_CORRELATION
      &END XC
      &POISSON
        POISSON_SOLVER MT
        PERIODIC NONE
      &END POISSON
      &PRINT
        &MO_CUBES
          FILENAME =HOMO.txt
          WRITE_CUBE FALSE
          NLUMO 10
          NHOMO 10
        &END MO_CUBES
      &END PRINT
    &END DFT
    &SUBSYS
      &CELL
        ABC 16.31807877 15.17847716 14.0643624
        PERIODIC NONE
      &END CELL
      &TOPOLOGY
        COORD_FILE_NAME dsgdb9nsd_123456.xyz
        COORD_FILE_FORMAT xyz
        &CENTER_COORDINATES
        &END CENTER_COORDINATES
      &END TOPOLOGY
      &KIND H
        RI_AUX_BASIS_SET aug-cc-pVDZ-RIFIT
        ELEMENT H
        POTENTIAL ALL
        BASIS_SET aug-cc-pVDZ
      &END KIND
      &KIND C
        RI_AUX_BASIS_SET aug-cc-pVDZ-RIFIT
        ELEMENT C
        POTENTIAL ALL
        BASIS_SET aug-cc-pVDZ
      &END KIND
      &KIND N
        RI_AUX_BASIS_SET aug-cc-pVDZ-RIFIT
        ELEMENT N
        POTENTIAL ALL
        BASIS_SET aug-cc-pVDZ
      &END KIND
      &KIND O
        RI_AUX_BASIS_SET aug-cc-pVDZ-RIFIT
        ELEMENT O
        POTENTIAL ALL
        BASIS_SET aug-cc-pVDZ
```

```
            &END KIND
            &KIND F
              RI_AUX_BASIS_SET aug-cc-pVDZ-RIFIT
              ELEMENT F
              POTENTIAL ALL
              BASIS_SET aug-cc-pVDZ
            &END KIND
            &KIND P
              RI_AUX_BASIS_SET aug-cc-pVDZ-RIFIT
              ELEMENT P
              POTENTIAL ALL
              BASIS_SET aug-cc-pVDZ
            &END KIND
            &KIND S
              RI_AUX_BASIS_SET aug-cc-pVDZ-RIFIT
              ELEMENT S
              POTENTIAL ALL
              BASIS_SET aug-cc-pVDZ
            &END KIND
            &KIND Cl
              RI_AUX_BASIS_SET aug-cc-pVDZ-RIFIT
              ELEMENT Cl
              POTENTIAL ALL
              BASIS_SET aug-cc-pVDZ
            &END KIND
            &KIND Br
              RI_AUX_BASIS_SET aug-cc-pVDZ-RIFIT
              ELEMENT Br
              POTENTIAL ALL
              BASIS_SET aug-cc-pVDZ
            &END KIND
            &KIND B
              RI_AUX_BASIS_SET aug-cc-pVDZ-RIFIT
              ELEMENT B
              POTENTIAL ALL
              BASIS_SET aug-cc-pVDZ
            &END KIND
            &KIND I
              RI_AUX_BASIS_SET aug-cc-pVDZ-RIFIT
              ELEMENT I
              POTENTIAL ALL
              BASIS_SET aug-cc-pVDZ
            &END KIND
        &END SUBSYS
    &END FORCE_EVAL
```

## Table 1. Hardware specifications

|   | CPU Model | Cluster |
|---|-----------|---------|
| 1 | Xeon(R) Gold 6252 CPU @ 2.10GHz | JUSTUS 2 |
| 2 | Intel(R) Xeon(R) CPU E5-2630 v3 @ 2.40GHz | MLS&WISO |
| 3 | Intel(R) Xeon(R) CPU E5-2640 v3 @ 2.60GHz | MLS&WISO |
| 4 | Intel(R) Xeon(R) CPU E5-4620 v3 @ 2.00GHz | MLS&WISO |
| 5 | AMD EPYC 7702 64-Core Processor | int-nano |
| 6 | AMD EPYC 7551P 32-Core Processor | int-nano |

# CPU time scaling

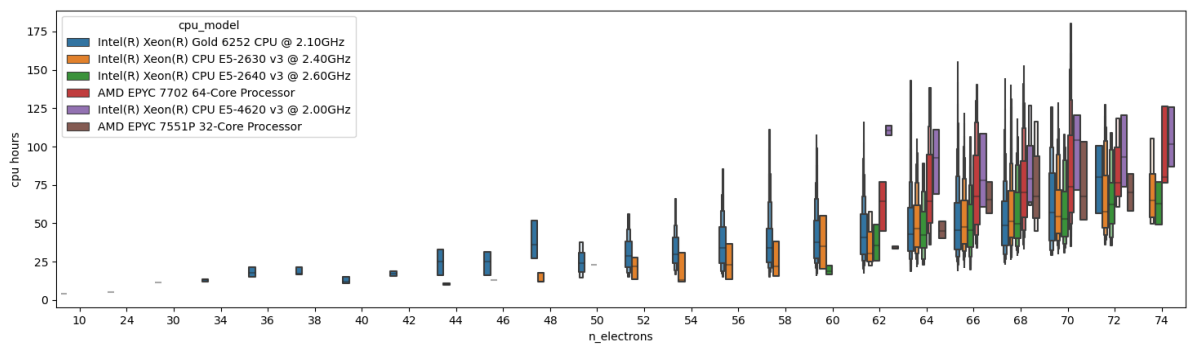

**Figure 1**. Computational time depending on the number of electrons split by different CPU model specifications. Total amount of computational resources in cpu hours: 7,439,925. This includes computing DFT electron density, which is used as a starting point for GW calculations, and GW calculations themselves. Both are done for two basis sets.